\documentclass[a4paper,english,prl,showpacs]{revtex4}
\usepackage[T1]{fontenc}
\usepackage{amsmath}
\usepackage[dvips]{graphicx}
\usepackage{amssymb}

\makeatletter
\usepackage{babel}
\makeatother
\begin{document}

\title{Universal Scaling Laws for Large Events in Driven Nonequilibrium
Systems}

\author{M. K. Verma}

\affiliation{Department of Physics, Indian Institute of Technology Kanpur, Kanpur
208016, India}

\author{S. Manna}

\affiliation{Department of Biophysics, University of Delhi South Campus, Benito
Juanez Road, New Delhi110021, India}

\author{J. Banerjee}

\affiliation{Department of Biophysics, University of Delhi South Campus, Benito
Juanez Road, New Delhi 110021, India}

\author{S. Ghosh}

\affiliation{Department of Animal Sciences, School of Life Sciences, University
of Hyderabad, Hyderabad 500046, India}

\begin{abstract}
For many driven-nonequilibrium systems, the probability distribution
functions of magnitude and recurrence-time of large events follow
a powerlaw indicating a strong temporal correlation. In this paper
we argue why these probability distribution functions are ubiquitous
in driven nonequilibrium systems, and we derive universal scaling
laws connecting the magnitudes, recurrence-time, and spatial intervals
of large events. The relationships between the scaling exponents have
also been studied. We show that the ion-channel current in Voltage-dependent
Anion Channels obeys the universal scaling law.
\end{abstract}


\maketitle
The physics of nonequilibrium systems has been of major interest in
last several decades. Recently large events have become a major resource
for characterizing nonequilibrium system. Bak et al. \cite{Bak:SOC_PRL,Bak:SOC_PRA}
constructed a model called {}``sandpile'' to abstract some of the
essential features of driven-nonequilibrium systems. Original model
of Bak et al. (BTW) as well as its variants exhibit powerlaw behaviour
for the magnitude and duration of avalanches, which play the role
of large events. Later scientists started investigating the temporal
correlations in various systems using recurrence-time pdf of large
events. In the BTW model, the pdf of recurrence time ($\tau$) of
large events is $\exp(-\tau/\tau_{c})/\tau_{c}$, which indicates
that the triggering of large events in this system are uncorrelated
\cite{Boff:SolarflarePRL,Sanc:SOCorNot}. This is a generic feature
of Poisson process. It was however soon realized that the events are
correlated in many nonequilibrium systems. Boffatta et al. \cite{Boff:SolarflarePRL}
studied solar flares, and found that the pdf of recurrence time for
the solar flares is a powerlaw, i. e., \begin{equation}
P(\tau)=A\tau^{-\beta},\label{eq:1}\end{equation}
 with $\beta\approx2.4$. These authors also studied the recurrence-time
distribution for the dissipation-bursts in magnetohydrodynamic turbulence,
and again found a powerlaw with $\beta\sim2.7$. These results show
that the large events in solar flares and magnetohydrodynamic turbulence
are temporally correlated. Note that the pdf of solar-flare magnitudes
is a powerlaw just like sandpile model \cite{Lu:Solarflare,Pacz:EQ_Intesity}.
The difference between the sandpile model and solar flare is in the
temporal correlations between the large events \cite{Boff:SolarflarePRL,Sanc:SOCorNot}.

Temporal correlations described above have been observed for many
other nonequilibrium systems. Corral \cite{Corr:PRL_Cluster,Corr:PRL_RG}
studied the pdf of recurrence time for large earthquakes (beyond 2.5
in Richter scale), and found it to be a powerlaw with $\beta\approx0.3$.
Bak et al. \cite{Bak:EQ} had done similar analysis. Earlier Guttenburg
and Richter \cite{GuteRich} showed that the magnitude of earthquakes
follows a powerlaw. {}``Extremal models'' (see Paczuski et al. \cite{Pacz:AvalacheDyna}
for review of these models) which are used to describe interface depinning,
biological evolution etc. also have similar temporal correlations.
Recently Banerjee et al. \cite{MKV:Ion} studied ion-channel currents
in Voltage Dependent Anion Channel (VDAC). Here too, the pdf of magnitude
as well as that of recurrence-time follow a powerlaw with $\beta\approx1.5$.
In thermal convection of Helium, Sreenivasan et al. \cite{KRS:Wind}
observed large-scale wind which switches directions; the pdf of inter-switch
interval is given by $(1/\tau)\exp(-\tau/\tau_{M})$. This feature
appears to indicate {}``release of convective stresses'' during
switching, just like energy release in earthquakes. The pdf of recurrence
time is connected to the power spectrum of the signal. Several research
groups \cite{Lowen:1f,Pacz:AvalacheDyna,Pacz:1f,MKV:Ion} have derived
$1/f$ power spectrum using the powerlaw distribution of recurrence-time.

There is a recent work on spatial correlations in driven nonequilibrium
systems. Davidsen and Paczuski \cite{Pacs:EQSpatial} computed the
correlation between the epicenters of subsequent earthquakes, and
found the pdf for the distance $\Delta r$ between the subsequent
epicenters to be a powerlaw. Earlier, Krishnamurthy and Barma \cite{Supr1,Supr2}
found a powerlaw distribution for the long-range jumps in self-organized
interface depinning model.

Corral \cite{Corr:PRL_Cluster,Corr:PRL_RG} and Bak et al. \cite{Bak:EQ}
combined the pdfs of magnitude and recurrence-time and derived a unified
scaling law. For events greater than a threshold $I_{c}$, then the
pdf for recurrence-time is

\begin{equation}
P(\tau)=Rf(\tau R)\label{eq:earthquake1}\end{equation}
with \begin{equation}
f(\tau R)\approx CR^{-0.33}\left(\frac{{\tau}}{a}\right)^{-0.33}\exp(-\tau R/a)\label{eq:earthquake2}\end{equation}
and $R=R_{0}10^{-bM_{c}}$($b\approx1$) is the average number of
earthquakes (with $I>I_{c}$) per unit time, and $a$ is a constant.ll
these work however are model specific. In our paper we present a general
scaling laws which should be applicable to all the above systems and
other driven nonequilibrium systems. Here we also derive universal
relationships between the exponents. Note that $M_{c}=\log_{10}A_{c}$
is the Richter scale. 

Eqs. (\ref{eq:earthquake1}, \ref{eq:earthquake2}) form the unified
scaling law for the earthquakes. One of the intriguing questions is
whether similar scaling laws exist for other nonequilibrium systems.
Another important puzzle is the origin of the unified scaling law.
In the present paper we address the origin of pdfs in driven nonequilibrium
systems, and we derive universal scaling laws connecting the magnitudes,
recurrence-time, and spatial intervals of large events. These scaling
laws should be applicable to all the above systems and other driven
nonequilibrium systems. Here we also derive universal relationships
between the scaling exponents. 

In the above mentioned driven-nonequilibrium system the large events
do not have a typical energy scale. Hence \emph{under steady-state,
the distribution of magnitude of large events follows a powerlaw.}
That is,\[
P(I)\sim I^{-\alpha},\]
where $I$ is the strength of the events, and $\alpha>0$. Note however
that the events cannot have energy more than what is available in
the system, hence the distribution of the tail is decaying, mostly
exponential. 

For temporal correlation study, we take events larger than a threshold
value $I_{c}$. Driven nonequilibrium systems can be classified in
two classes, ones in which the events are uncorrelated, and the others
in which the events are correlated. In the uncorrelated case, the
pdf for recurrence of large events is Poissonian because the probability
of large event is small. If the large events occur with certain rate
$R$, then $1/R$ provides the time-scale for the recurrence of large
events; consequently the pdf will be $P(\tau)\sim R\exp(-R\tau)$.
The above feature is observed in BTW model \cite{Sanc:SOCorNot}.
In the correlated case, the nonequilibrium system is {}``\emph{critica}l''
under steady-state, and the system does not have any time-scale for
the recurrence time. Hence, the pdf of recurrence time is a powerlaw.
We can combine the pdf of magnitude and recurrence time using the
following law:\begin{equation}
P(I_{c},\tau)=I_{c}^{-\alpha}\tau^{-\beta}f(-\tau/\tau_{c}),\label{eq:Ptau}\end{equation}
where $f(x)$ is an universal function. For small $x$, $f(x)=1$,
and for large $x$, $f(x)\rightarrow\exp(-x)$. Further application
of scaling yields $\tau_{c}\sim I_{c}^{\gamma}$. It is expected that
when $I_{c1}>I_{c2}$, $P(I_{c1},\tau)<P(I_{c2},\tau)$, but $\tau_{c1}>\tau_{c2}$
\cite{Corr:PRL_Cluster,Corr:PhysicaA}. This observation implies that
$\gamma>0$. Corral's law for earthquake \cite{Corr:PRL_Cluster,Corr:PRL_RG}
is a good example which illustrates the above argument.

The exponents $\alpha,\beta$, and $\gamma$ are not all independent.
Using conservation of probability\begin{equation}
\int P(I_{c},\tau)d\tau=1,\label{eq:P-conserve}\end{equation}
and equating the powers of $I_{c}$ on both sides, we obtain

\begin{equation}
\alpha=\gamma(1-\beta).\label{eq:reln1}\end{equation}
As an example, the earthquake exponents of Corral \cite{Corr:PRL_Cluster,Corr:PhysicaA}
are $\alpha\approx2/3$, $\gamma\approx1$, $\beta\approx0.3$, which
satisfy the above relationship to a good approximation.

Unfortunately Eq. (\ref{eq:P-conserve}) does not converge for $P(I_{c},\tau)$
with $\beta\ge1$ {[}Eq. (\ref{eq:Ptau}){]}. However we can circumvent
this difficulty by computing the higher-order moments of $\tau$.
If the integer part of $\beta$ is $n$, then we compute\begin{equation}
\left\langle \tau^{n}\right\rangle =\int\tau^{n}P(I_{c},\tau)d\tau.\label{eq:tau_n}\end{equation}
Clearly the above integral converges. By postulating $\left\langle \tau^{n}\right\rangle \sim I_{c}^{\delta_{n}}$,
and equating the exponents of $I_{c}$ on both sides, we obtain

\begin{equation}
\alpha=\gamma(n+1-\beta)-\delta_{n}.\label{eq:reln2}\end{equation}
Hence the scaling arguments provide us an expression for $P(I_{c},\tau)$
{[}Eq. (\ref{eq:Ptau}){]} and relationships between exponents.

The form of pdf of spatial separations between consecutive large events
can be derived in the similar lines. The events could be either spatially
uncorrelated or correlated. For uncorrelated case, the pdf of spatial
separation is expected to be Poissonian: $P(\Delta r)\sim\exp(-\Delta r/(\Delta r)_{c})/(\Delta r)_{c}$,
where the spatial density of the large events provide a length-scale
$(\Delta r)_{c}$. Nonequilibrium systems with spatially correlated
events are \emph{critical} under steady-state, and they do not have
any length-scale for the separation between the consecutive large
events. For this class of systems, the pdf is \cite{Pacs:EQSpatial}\begin{equation}
P(I_{c},\Delta r)=I_{c}^{-\alpha}(\Delta r)^{-\zeta}\exp(-\Delta r/(\Delta r)_{c}),\label{eq:Pdr}\end{equation}
with $(\Delta r)_{c}\sim I_{c}^{-\mu}$. If we assume dynamical scaling
$\tau\sim(\Delta r)^{z}$, then $P(I_{c},\Delta r)=P(I_{c},\tau)d\tau/d(\Delta r)$
yields \begin{equation}
\zeta=z(\beta-1)+1.\label{eq:spacetime}\end{equation}
Currently we do not know all the above parameters for any driven nonequilibrium
systems. It will be interesting to test this conjecture.

It should be noted that earlier Maslov et al. \cite{Pacz:1f}, Paczuski
et al. \cite{Pacz:AvalacheDyna}, and others had derived relationships
between scaling exponents for several driven nonequilibrium systems,
e. g., Bak-Sneppen model, Directed Percolation etc. \cite{Pacz:AvalacheDyna}.
These relationships differ significantly from our equations, which
are based on large events. 

The scaling laws discussed above are quite general, and are expected
to hold for systems with correlated large events. Earthquake is one
such system for which Corral \cite{Corr:PRL_Cluster,Corr:PhysicaA}
, Bak et al. \cite{Bak:EQ}, and Davidsen and Paczuski \cite{Pacs:EQSpatial}
have computed some of the scaling exponents. Solar flares, Bak-Sneppen
model for punctuated evolution, and turbulence are other nonequilibrium
systems which too have similar behaviour. \emph{It should be noted
however that the exponents vary for different systems, but the form
of pdfs is the same for all them.} In the following discussion we
apply the above scaling ansatz to ion-channel currents in voltage-dependent
anion channels (VDAC), and test if it works. 

Our experimental setup on ion-channel current consists of two chambers
containing buffer solutions \emph{KCl}, $MgCl_{2}$, and HEPES. The
chambers are separated by a thin wall of a polystyrene cuvette. A
matched pair of \emph{Ag/AgCl} electrodes, connected to Axopatch-200
amplifier, are put into the buffer solution of two chambers. A bilayer
lipid membrane is formed on the pore. The protein molecules of VDAC
purified from rat brain mitochondria (De Pinto et al. \cite{Pinto:ion})
are put into the buffer solution of the outer chamber, after which
they are stirred magnetically. The VDAC molecules now reconstitute
the lipid membrane, and a channel is formed (Roos et al. \cite{Roos:ion}).
After one channel is formed, the buffer solution (containing more
protein molecules) of the outer chamber is replaced by fresh buffer
solution. Thus we form a single ion-channel. 

\begin{figure}
\includegraphics{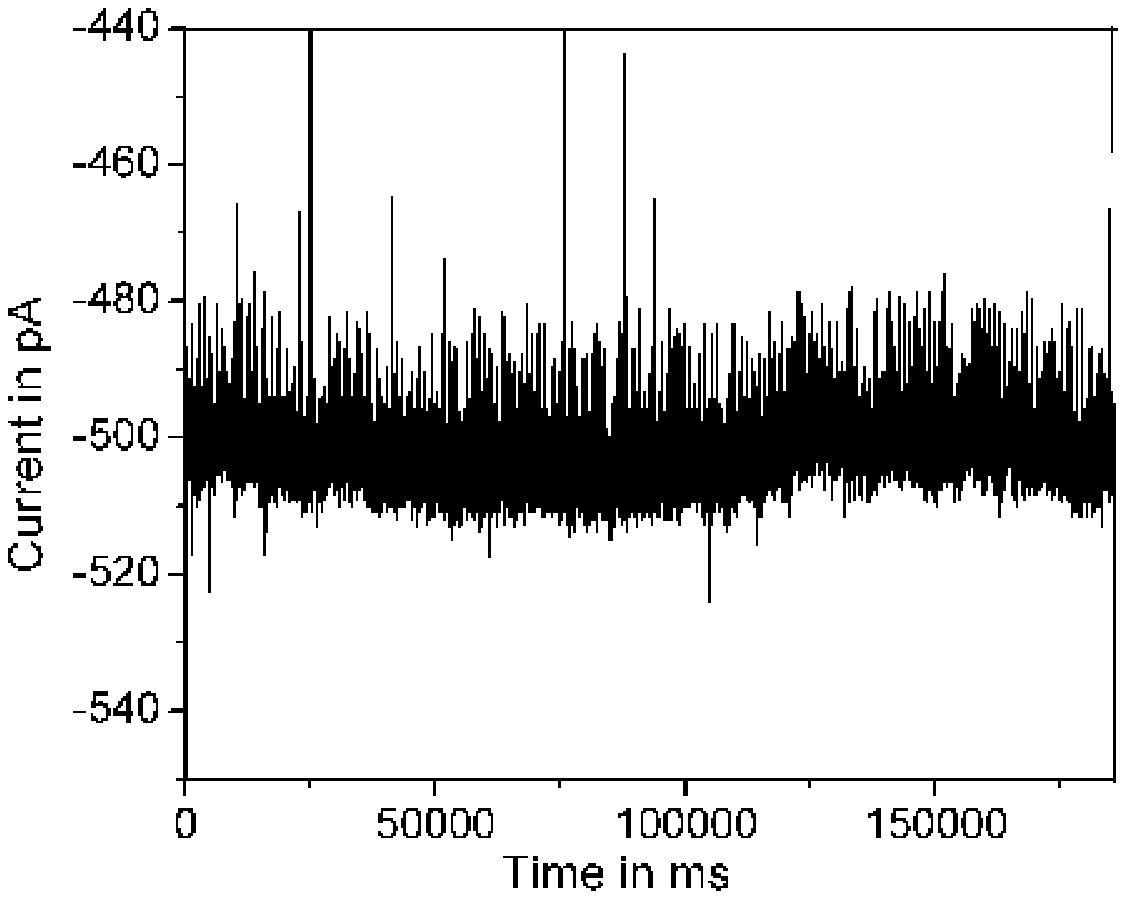}

\caption{\label{Fig:timeseries} Timeseries of ion-channel currents in open
state of Voltage-dependent Anion Channels (VDAC). Four parts have
been patched here for extending the data range.}
\end{figure}

Now the voltage is applied across the membrane, and the resulting
current is recorded using the data acquisition software Clampex. This
method is same as that used by Bezrukov et al. \cite{Bezr:Ion} and
Banerjee et al \cite{MKV:Ion}. In our experiment we measure currents
with sampling frequency 1000 Hz
 for various applied voltages in the range $-30mV$ to $30mV$. Typically
the current has open and closed states with a fractal structure \cite{Lieb:ion}.
We find the length of a single open state to be too small. Therefore,
we patch four long open-states and use the resulting timeseries for
our analysis (see Fig. \ref{Fig:timeseries}). Due to a large variation
of currents in various patches we perform the running average of the
timeseries over eleven neighboring points. The noise is obtained by
filtering the running average from the time-series. The statistics
is performed on the resulting noise time-series.

\begin{figure}
\includegraphics{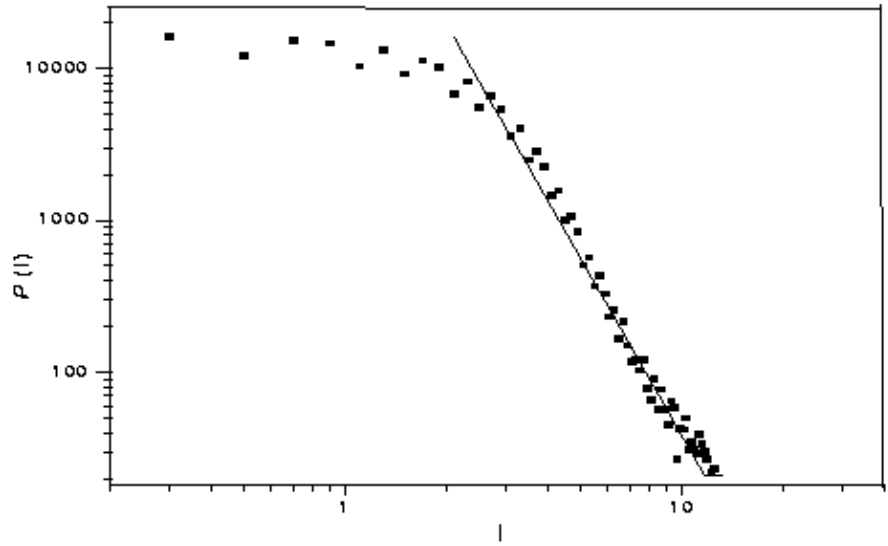}

\caption{\label{Fig:PIvsI} Plot of the probability distribution of magnitudes
of current fluctuation $P(I)$ in ion-channel. The best fit is a powerlaw
with slope $\alpha=3.86\pm0.08$.}
\end{figure}

We take the absolute value of the current fluctuations, and compute
the pdf of the magnitude of current fluctuation $P(I)$. The log-log
plot of $P(I)$ vs. $I$ is shown in Fig. \ref{Fig:PIvsI}. The slope
of the plot provides us the exponent $\alpha=3.86\pm0.08$.

\begin{figure}
\includegraphics[%
  scale=0.5]{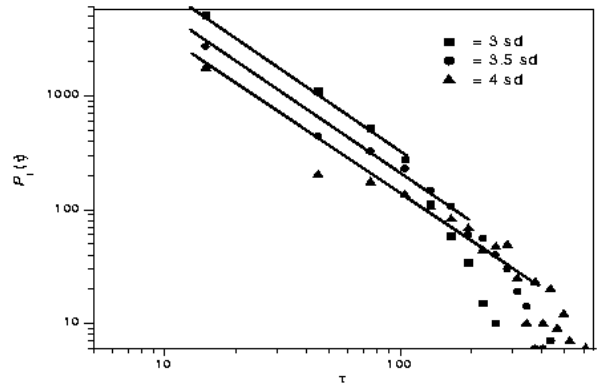}

\caption{\label{Fig:Ptvst} Plot of the probability distribution of recurrence
time $P(I_{c},\tau)$for three different threshold currents $I_{c}=3\sigma,3.5\sigma,4\sigma$,
where $\sigma$ is the standard deviations. The best fit for all the
three plots is a powerlaw with the exponent $\beta=1.47\pm0.05$. }
\end{figure}

For the computation of temporal correlation, we filter the current
using a threshold value $I_{c}$. From the filtered current, we compute
time-interval between subsequent current peaks. From this set of time-interval
we compute the pdf of recurrence time $P(I_{c},\tau)$. We perform
this exercise for three values of $I_{c}:3\sigma,3.5\sigma,4\sigma$,
where $\sigma$ is the standard deviation of the current fluctuations.
The corresponding plots $P(I_{c},\tau)$ vs. $\tau$ are shown in
Fig. \ref{Fig:Ptvst}. All the three plots have approximately the
same exponent, $\beta=1.47\pm0.05$. Note that for higher $I_{c}$,
$P(I_{c},\tau)$ is lower, but the powerlaw extends up to higher $\tau$.
This feature is also seen in earthquakes.

\begin{figure}
\includegraphics[%
  scale=0.8]{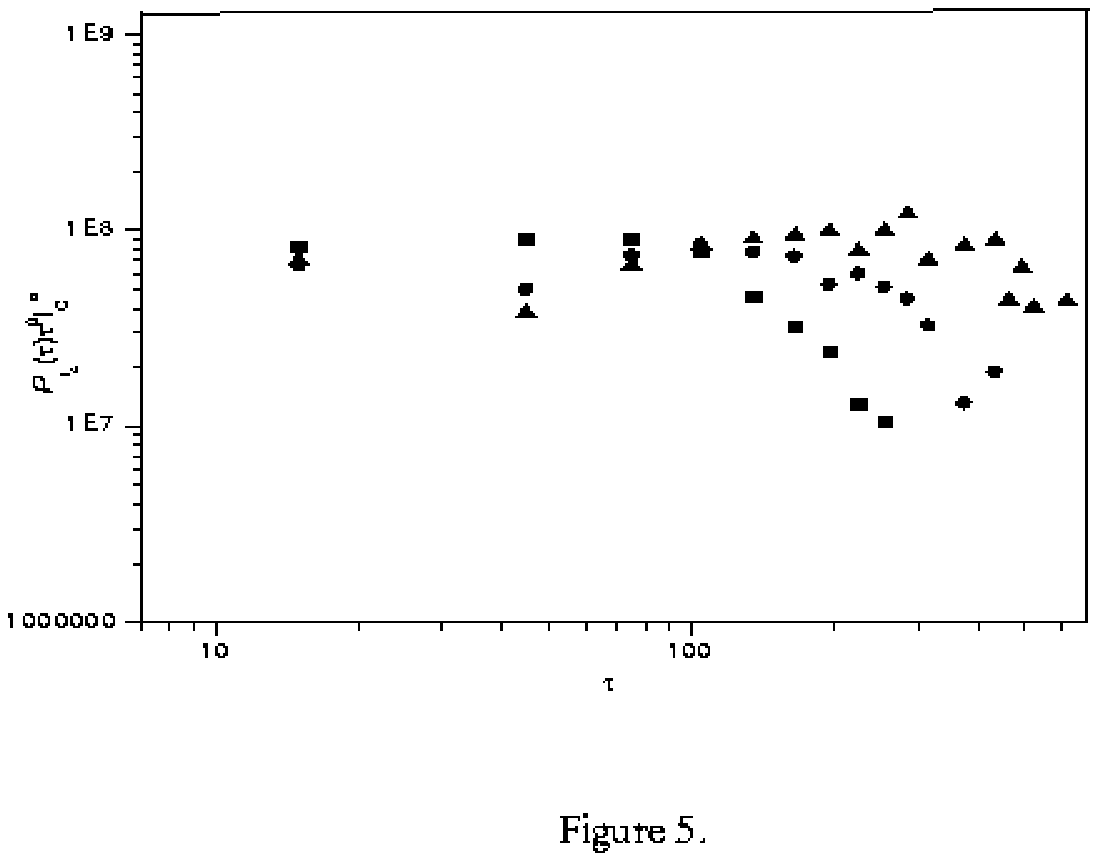}

\caption{\label{Fig:norm1} Plot of $P(I_{c},\tau)\tau^{\beta}$ vs. $\tau$.}
\end{figure}

\begin{figure}
\includegraphics[%
  scale=0.8]{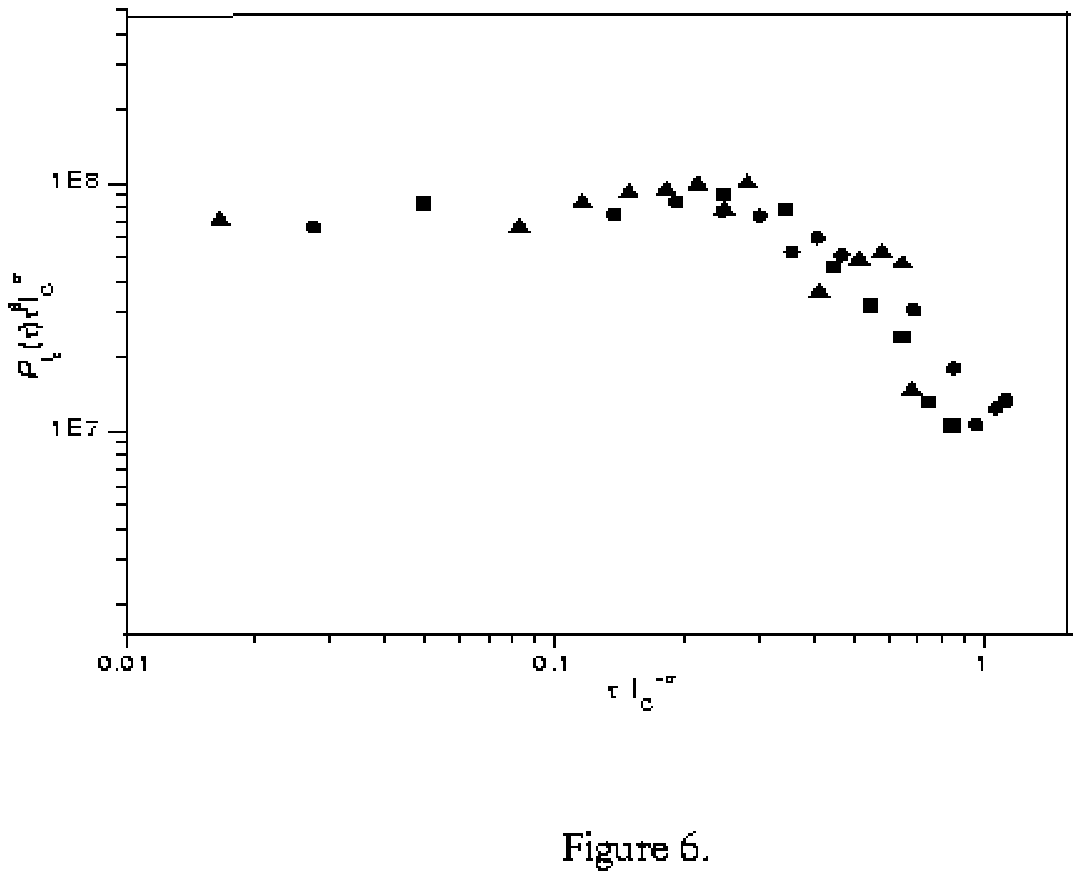}

\caption{\label{Fig:norm2} Plot of $P(I_{c},\tau)\tau^{\beta}I_{c}^{\alpha}$
vs. $\tau I_{c}^{-\alpha}$. The resulting plot is the universal function
$f(x)$.}
\end{figure}

To obtain the universal function, we need to scale both $x$ and $y$
axes. If we scale only the $y$-axis to $P(I,\tau)I_{c}^{\alpha}\tau^{\beta}$
and plot of $P(I,\tau)I_{c}^{\alpha}\tau^{\beta}$ vs. $\tau$, then
we obtain a reasonable compactification along $y$ axis, but not along
$x$ axis. See Fig. \ref{Fig:norm1} for illustration. When we scale
$x$-axis by $\tau_{c}=I_{c}^{\gamma}$ and plot $P(I,\tau)I_{c}^{\alpha}\tau^{\beta}$vs.
$\tau I_{c}^{-\alpha}$ (see Fig. \ref{Fig:norm2}), the compactification
is quite good. The resulting plot is the universal function $f(x)$
of Eq. (\ref{eq:Ptau}). Note that we have taken $\gamma=\alpha$,
which is an approximation. The actual parameters $\gamma$ and $\delta_{1}$
{[}Eq. (\ref{eq:reln2}){]} are not accessible in the present experiment.
Since the current flow in the ion-channels is a localized, we do not
have the pdf for spatial separation between consecutive events. 

In summary, the pdf of magnitudes of events in a driven nonequilibrium
systems follow a powerlaw due to lack of any typical energy scale
for the events. The pdf of recurrence-time could be either an exponential
or a powerlaw depending on whether the events are temporally uncorrelated
or correlated. The pdf of recurrence time for the many correlated
nonequilibrium systems is a powerlaw due to lack of any timescale
in the system. When the pdfs of magnitudes and recurrence-time for
the large events are powerlaws, they could be combined into a single
universal scaling law. In the present paper we derive the universal
scaling law as well as relationships between the scaling exponents
for driven-nonequilibrium systems. We show that universal scaling
law is satisfied by ion-channel currents. Earlier, Corral \cite{Corr:PRL_Cluster,Corr:PRL_RG}
and Bak et al. \cite{Bak:EQ} had derived similar law for earthquakes.
We have also derived similar scaling law for the pdf of spatial separation
between subsequent large events. We should however keep in mind that
the exponents vary for different systems, but the form of pdfs is
the same for all them. \emph{}

Temporal and spatial correlations between events are seen in solar
flares, turbulence, interface depinning models, music, punctuated
equilibria etc. A timeseries of major historical events would presumably
have temporal and spatial correlations. We should check how well scaling
laws describe the above correlations. Higher-order correlations between
the events should be another source for understanding nonequilibrium
systems.

\begin{acknowledgments}
MKV acknowledges the kind hospitality of D. Ghosh, his group, and
his mother during his visits to Delhi University. MKV also thanks
Harshawardhan Wanare, A. Corral, Amit Dutta, Sutapa Mukherjee, Archisman
Ghosh, Amar Chandra, and Mustansir Barma for useful discussions.
\end{acknowledgments}
\bibliographystyle{apsrev}

\begin{thebibliography}{23}
\expandafter\ifx\csname natexlab\endcsname\relax\def\natexlab#1{#1}\fi
\expandafter\ifx\csname bibnamefont\endcsname\relax
  \def\bibnamefont#1{#1}\fi
\expandafter\ifx\csname bibfnamefont\endcsname\relax
  \def\bibfnamefont#1{#1}\fi
\expandafter\ifx\csname citenamefont\endcsname\relax
  \def\citenamefont#1{#1}\fi
\expandafter\ifx\csname url\endcsname\relax
  \def\url#1{\texttt{#1}}\fi
\expandafter\ifx\csname urlprefix\endcsname\relax\def\urlprefix{URL }\fi
\providecommand{\bibinfo}[2]{#2}
\providecommand{\eprint}[2][]{\url{#2}}

\bibitem[{\citenamefont{Bak et~al.}(1987)\citenamefont{Bak, Tang, and
  Wiesenfeld}}]{Bak:SOC_PRL}
\bibinfo{author}{\bibfnamefont{P.}~\bibnamefont{Bak}},
  \bibinfo{author}{\bibfnamefont{C.}~\bibnamefont{Tang}}, \bibnamefont{and}
  \bibinfo{author}{\bibfnamefont{K.}~\bibnamefont{Wiesenfeld}},
  \bibinfo{journal}{Phys. Rev. Lett.} \textbf{\bibinfo{volume}{59}},
  \bibinfo{pages}{381} (\bibinfo{year}{1987}).

\bibitem[{\citenamefont{Bak et~al.}(1988)\citenamefont{Bak, Tang, and
  Wiesenfeld}}]{Bak:SOC_PRA}
\bibinfo{author}{\bibfnamefont{P.}~\bibnamefont{Bak}},
  \bibinfo{author}{\bibfnamefont{C.}~\bibnamefont{Tang}}, \bibnamefont{and}
  \bibinfo{author}{\bibfnamefont{K.}~\bibnamefont{Wiesenfeld}},
  \bibinfo{journal}{Phys. Rev. A} \textbf{\bibinfo{volume}{38}},
  \bibinfo{pages}{364} (\bibinfo{year}{1988}).

\bibitem[{\citenamefont{Boffetta et~al.}(1999)\citenamefont{Boffetta, Carbone,
  and G}}]{Boff:SolarflarePRL}
\bibinfo{author}{\bibfnamefont{G.}~\bibnamefont{Boffetta}},
  \bibinfo{author}{\bibfnamefont{V.}~\bibnamefont{Carbone}}, \bibnamefont{and}
  \bibinfo{author}{\bibfnamefont{P.}~\bibnamefont{G}}, \bibinfo{journal}{Phys.
  Rev. Lett.} \textbf{\bibinfo{volume}{83}}, \bibinfo{pages}{4662}
  (\bibinfo{year}{1999}).

\bibitem[{\citenamefont{S\'anchez et~al.}(2002)\citenamefont{S\'anchez, Newman,
  and Carreeras}}]{Sanc:SOCorNot}
\bibinfo{author}{\bibfnamefont{R.}~\bibnamefont{S\'anchez}},
  \bibinfo{author}{\bibfnamefont{D.~E.} \bibnamefont{Newman}},
  \bibnamefont{and} \bibinfo{author}{\bibfnamefont{B.~A.}
  \bibnamefont{Carreeras}}, \bibinfo{journal}{Phys. Rev. Lett.}
  \textbf{\bibinfo{volume}{88}}, \bibinfo{pages}{068302}
  (\bibinfo{year}{2002}).

\bibitem[{\citenamefont{Lu and Hamilton}(1991)}]{Lu:Solarflare}
\bibinfo{author}{\bibfnamefont{E.~T.} \bibnamefont{Lu}} \bibnamefont{and}
  \bibinfo{author}{\bibfnamefont{R.~J.} \bibnamefont{Hamilton}},
  \bibinfo{journal}{Astrophys. J.} \textbf{\bibinfo{volume}{380}},
  \bibinfo{pages}{L89} (\bibinfo{year}{1991}).

\bibitem[{\citenamefont{Baiesi et~al.}(2006)\citenamefont{Baiesi, Paczuski, and
  Stella}}]{Pacz:EQ_Intesity}
\bibinfo{author}{\bibfnamefont{M.}~\bibnamefont{Baiesi}},
  \bibinfo{author}{\bibfnamefont{M.}~\bibnamefont{Paczuski}}, \bibnamefont{and}
  \bibinfo{author}{\bibfnamefont{A.~L.} \bibnamefont{Stella}},
  \bibinfo{journal}{Phys. Rev. Lett.} \textbf{\bibinfo{volume}{96}},
  \bibinfo{pages}{051103} (\bibinfo{year}{2006}).

\bibitem[{\citenamefont{Corral}(2004{\natexlab{a}})}]{Corr:PRL_Cluster}
\bibinfo{author}{\bibfnamefont{A.}~\bibnamefont{Corral}},
  \bibinfo{journal}{Phys. Rev. Lett.} \textbf{\bibinfo{volume}{92}},
  \bibinfo{pages}{108501} (\bibinfo{year}{2004}{\natexlab{a}}).

\bibitem[{\citenamefont{Corral}(2005)}]{Corr:PRL_RG}
\bibinfo{author}{\bibfnamefont{A.}~\bibnamefont{Corral}},
  \bibinfo{journal}{Phys. Rev. Lett.} \textbf{\bibinfo{volume}{95}},
  \bibinfo{pages}{028501} (\bibinfo{year}{2005}).

\bibitem[{\citenamefont{Bak et~al.}(2002)\citenamefont{Bak, Christensen, Danon,
  and Scanlon}}]{Bak:EQ}
\bibinfo{author}{\bibfnamefont{P.}~\bibnamefont{Bak}},
  \bibinfo{author}{\bibfnamefont{K.}~\bibnamefont{Christensen}},
  \bibinfo{author}{\bibfnamefont{L.}~\bibnamefont{Danon}}, \bibnamefont{and}
  \bibinfo{author}{\bibfnamefont{T.}~\bibnamefont{Scanlon}},
  \bibinfo{journal}{Phys. Rev. Lett.} \textbf{\bibinfo{volume}{88}},
  \bibinfo{pages}{178501} (\bibinfo{year}{2002}).

\bibitem[{\citenamefont{Gutenberg and Richter}(1936)}]{GuteRich}
\bibinfo{author}{\bibfnamefont{B.}~\bibnamefont{Gutenberg}} \bibnamefont{and}
  \bibinfo{author}{\bibfnamefont{C.~F.} \bibnamefont{Richter}},
  \bibinfo{journal}{Science} \textbf{\bibinfo{volume}{83}},
  \bibinfo{pages}{183} (\bibinfo{year}{1936}).

\bibitem[{\citenamefont{Paczuski et~al.}(1996)\citenamefont{Paczuski, Maslov,
  and Bak}}]{Pacz:AvalacheDyna}
\bibinfo{author}{\bibfnamefont{M.}~\bibnamefont{Paczuski}},
  \bibinfo{author}{\bibfnamefont{S.}~\bibnamefont{Maslov}}, \bibnamefont{and}
  \bibinfo{author}{\bibfnamefont{P.}~\bibnamefont{Bak}},
  \bibinfo{journal}{Phys. Rev. E} \textbf{\bibinfo{volume}{53}},
  \bibinfo{pages}{414} (\bibinfo{year}{1996}).

\bibitem[{\citenamefont{Banerjee et~al.}(2006)\citenamefont{Banerjee, Verma,
  Manna, and Ghosh}}]{MKV:Ion}
\bibinfo{author}{\bibfnamefont{J.}~\bibnamefont{Banerjee}},
  \bibinfo{author}{\bibfnamefont{M.~K.} \bibnamefont{Verma}},
  \bibinfo{author}{\bibfnamefont{S.}~\bibnamefont{Manna}}, \bibnamefont{and}
  \bibinfo{author}{\bibfnamefont{S.}~\bibnamefont{Ghosh}},
  \bibinfo{journal}{Europhys. Lett.} \textbf{\bibinfo{volume}{73}},
  \bibinfo{pages}{457} (\bibinfo{year}{2006}).

\bibitem[{\citenamefont{Sreenivasan et~al.}(2002)\citenamefont{Sreenivasan,
  Bershadskii, and Niemela}}]{KRS:Wind}
\bibinfo{author}{\bibfnamefont{K.~R.} \bibnamefont{Sreenivasan}},
  \bibinfo{author}{\bibfnamefont{A.}~\bibnamefont{Bershadskii}},
  \bibnamefont{and} \bibinfo{author}{\bibfnamefont{J.~J.}
  \bibnamefont{Niemela}}, \bibinfo{journal}{Phys. Rev. E}
  \textbf{\bibinfo{volume}{65}}, \bibinfo{pages}{56306} (\bibinfo{year}{2002}).

\bibitem[{\citenamefont{Lowen and Teich}(1993)}]{Lowen:1f}
\bibinfo{author}{\bibfnamefont{S.~B.} \bibnamefont{Lowen}} \bibnamefont{and}
  \bibinfo{author}{\bibfnamefont{M.~C.} \bibnamefont{Teich}},
  \bibinfo{journal}{Phys. Rev. E} \textbf{\bibinfo{volume}{47}},
  \bibinfo{pages}{992} (\bibinfo{year}{1993}).

\bibitem[{\citenamefont{Maslow et~al.}(1994)\citenamefont{Maslow, Paczuski, and
  Bak}}]{Pacz:1f}
\bibinfo{author}{\bibfnamefont{S.}~\bibnamefont{Maslow}},
  \bibinfo{author}{\bibfnamefont{M.}~\bibnamefont{Paczuski}}, \bibnamefont{and}
  \bibinfo{author}{\bibfnamefont{P.}~\bibnamefont{Bak}},
  \bibinfo{journal}{Phys. Rev. Lett.} \textbf{\bibinfo{volume}{73}},
  \bibinfo{pages}{2162} (\bibinfo{year}{1994}).

\bibitem[{\citenamefont{Davidsen and Paczuski}(2005)}]{Pacs:EQSpatial}
\bibinfo{author}{\bibfnamefont{J.}~\bibnamefont{Davidsen}} \bibnamefont{and}
  \bibinfo{author}{\bibfnamefont{M.}~\bibnamefont{Paczuski}},
  \bibinfo{journal}{Phys. Rev. Lett.} \textbf{\bibinfo{volume}{94}},
  \bibinfo{pages}{048501} (\bibinfo{year}{2005}).

\bibitem[{\citenamefont{Krishnamurthy and Barma}(1996)}]{Supr1}
\bibinfo{author}{\bibfnamefont{S.}~\bibnamefont{Krishnamurthy}}
  \bibnamefont{and} \bibinfo{author}{\bibfnamefont{M.}~\bibnamefont{Barma}},
  \bibinfo{journal}{Phys. Rev. Lett.} \textbf{\bibinfo{volume}{76}},
  \bibinfo{pages}{423} (\bibinfo{year}{1996}).

\bibitem[{\citenamefont{Krishnamurthy and Barma}(1998)}]{Supr2}
\bibinfo{author}{\bibfnamefont{S.}~\bibnamefont{Krishnamurthy}}
  \bibnamefont{and} \bibinfo{author}{\bibfnamefont{M.}~\bibnamefont{Barma}},
  \bibinfo{journal}{Phys. Rev. E} \textbf{\bibinfo{volume}{57}},
  \bibinfo{pages}{2949} (\bibinfo{year}{1998}).

\bibitem[{\citenamefont{Corral}(2004{\natexlab{b}})}]{Corr:PhysicaA}
\bibinfo{author}{\bibfnamefont{A.}~\bibnamefont{Corral}},
  \bibinfo{journal}{Physica A} \textbf{\bibinfo{volume}{340}},
  \bibinfo{pages}{590} (\bibinfo{year}{2004}{\natexlab{b}}).

\bibitem[{\citenamefont{De~Pinto et~al.}(1987)\citenamefont{De~Pinto, Prezioso,
  and Palmieri}}]{Pinto:ion}
\bibinfo{author}{\bibfnamefont{V.}~\bibnamefont{De~Pinto}},
  \bibinfo{author}{\bibfnamefont{G.}~\bibnamefont{Prezioso}}, \bibnamefont{and}
  \bibinfo{author}{\bibfnamefont{F.}~\bibnamefont{Palmieri}},
  \bibinfo{journal}{Biochim. Biophys. Acta} \textbf{\bibinfo{volume}{905}},
  \bibinfo{pages}{499} (\bibinfo{year}{1987}).

\bibitem[{\citenamefont{Roos et~al.}(1982)\citenamefont{Roos, Benz, and
  Brdiczka}}]{Roos:ion}
\bibinfo{author}{\bibfnamefont{N.}~\bibnamefont{Roos}},
  \bibinfo{author}{\bibfnamefont{R.}~\bibnamefont{Benz}}, \bibnamefont{and}
  \bibinfo{author}{\bibfnamefont{D.}~\bibnamefont{Brdiczka}},
  \bibinfo{journal}{Biochim. Biophys. Acta} \textbf{\bibinfo{volume}{686}},
  \bibinfo{pages}{204} (\bibinfo{year}{1982}).

\bibitem[{\citenamefont{Bezrukov and Winterhalter}(2000)}]{Bezr:Ion}
\bibinfo{author}{\bibfnamefont{S.~M.} \bibnamefont{Bezrukov}} \bibnamefont{and}
  \bibinfo{author}{\bibfnamefont{M.}~\bibnamefont{Winterhalter}},
  \bibinfo{journal}{Phys. Rev. Lett.} \textbf{\bibinfo{volume}{85}},
  \bibinfo{pages}{202} (\bibinfo{year}{2000}).

\bibitem[{\citenamefont{Liebovitch et~al.}(2001)\citenamefont{Liebovitch,
  Scheurle, Rusek, and Zochowski}}]{Lieb:ion}
\bibinfo{author}{\bibfnamefont{L.~S.} \bibnamefont{Liebovitch}},
  \bibinfo{author}{\bibfnamefont{D.}~\bibnamefont{Scheurle}},
  \bibinfo{author}{\bibfnamefont{M.}~\bibnamefont{Rusek}}, \bibnamefont{and}
  \bibinfo{author}{\bibfnamefont{M.}~\bibnamefont{Zochowski}},
  \bibinfo{journal}{Methods} \textbf{\bibinfo{volume}{24}},
  \bibinfo{pages}{359} (\bibinfo{year}{2001}).

\end{thebibliography}

\begin{center}\textbf{\Large Figure Caption}\end{center}{\Large \par}

Figure 1: Timeseries of ion-channel currents in open state of Voltage-dependent
Anion Channels (VDAC). Four parts have been patched here for extending
the data range.

Figure 2: Plot of the probability distribution of magnitudes of current
fluctuation in ion-channel. The best fit is a powerlaw with slope
$\alpha=3.86\pm0.08$.

Figure 3: Plot of the probability distribution of recurrence time
$P(I_{c},\tau)$ for three different threshold currents $I_{c}=3\sigma,3.5\sigma,4\sigma$,
where $\sigma$ is the standard deviations. The best fit for all the
three plots is a powerlaw with the exponent $\beta=1.47\pm0.05$.

Figure 4: Plot of $P(I_{c},\tau)\tau^{\beta}$ vs. $\tau$.

Figure 5: Plot of $P(I_{c},\tau)\tau^{\beta}I_{c}^{\alpha}$ vs. $\tau I_{c}^{-\alpha}$.
The resulting plot is the universal function $f(x)$.

\pagebreak

\begin{center}\textbf{\Large Figures}\end{center}{\Large \par}

\pagebreak

\pagebreak

\pagebreak

\end{document}